# Drastic suppression of superconducting $T_c$ by anisotropic strain near a nematic quantum critical point

Paul Malinowski[1], Qianni Jiang[1], Joshua Sanchez[1], Zhaoyu Liu[1], Joshua Mutch[1], Preston Went[1], Jian Liu[2], Philip Ryan[3,4], Jong-Woo Kim[3], Jiun-Haw Chu[1*].

**Affiliations:**

[1]Department of Physics, University of Washington, Seattle, Washington 98195, USA.

[2]Department of Physics, University of Tennessee, Knoxville, Tennessee 37996, USA.

[3]Advanced Photon Source, Argonne National Laboratories, Lemont, Illinois 60439, USA.

[4]School of Physical Sciences, Dublin City University, Dublin 9, Ireland.

*Correspondence to: jhchu@uw.edu (J.-H.C)

**Abstract:**

High temperature superconductivity emerges in the vicinity of competing strongly correlated phases. In the iron-based superconductor Ba(Fe$_{1-x}$Co$_x$)$_2$As$_2$, the superconducting state shares the composition-temperature phase diagram with an electronic nematic phase and an antiferromagnetic phase that break the crystalline rotational symmetry. Symmetry considerations suggest that anisotropic strain can enhance these competing phases and thus suppress the superconductivity. Here we study the effect of anisotropic strain on the superconducting transition in single crystals of Ba(Fe$_{1-x}$Co$_x$)$_2$As$_2$ through electrical transport, magnetic susceptibility, and x-ray diffraction measurements. We find that in the underdoped and near-optimally doped regions of the phase diagram, the superconducting critical temperature is rapidly suppressed by both compressive and tensile stress, and in the underdoped case this suppression is enough to induce a strain-tuned superconductor to metal quantum phase transition.

**Main Text:**

Most unconventional and high temperature superconductors share a similar phase diagram. As the system is tuned by chemical doping or pressure, superconductivity emerges as a nearby symmetry breaking phase is suppressed. Such a phase diagram has been observed in the antiferromagnetic heavy fermion compounds ([1], [2]), the charge density wave transition metal dichalcogenides ([3], [4]), the spin density wave quasi-1D organics ([5], [6]), and the iron-based superconductors ([7], [8]). The empirical observation of this common phase diagram has led to the belief that the symmetry breaking phase adjacent to unconventional superconductivity plays a dual role: its static order competes with superconductivity, yet its fluctuations are beneficial, if not responsible, for the superconducting pairing ([9], [10]).

While there is a large body of experimental work that supports this long-held view, most studies rely on tracking the properties of materials as a function of position in the phase diagram. Another simple way to verify the above scenario is applying the symmetry breaking field conjugate to the competing phase at a fixed position in the phase diagram. The conjugate field enhances the static order and suppresses the dynamic fluctuations, which should strongly suppress the superconducting $T_c$. The demonstration of this field-controlled $T_c$ is not only a direct verification of the hypothesis discussed above, but also has profound implications for the technological applications of unconventional superconductors. Nevertheless, most of the competing phases in unconventional superconductors break translational symmetry, which requires a spatially modulated conjugate field that is difficult to realize experimentally.

The electronic nematic phase ([11]) in the iron-based superconductors is a rare exception, only breaking the crystalline fourfold rotational symmetry. In the iron pnictides a collinear antiferromagnetic order further breaks translational and time reversal symmetries within the nematic phase (*12*, *13*), whereas in the iron chalcogenides nematic order may exist without long-range magnetic order (*14*, *15*). Due to the finite electron-lattice coupling, the nematicity induces an orthorhombic structural distortion that can be considered as a secondary order parameter. X-ray diffraction measurements have shown a suppression of the orthorhombic distortion upon entering the superconducting state (*16*), which is direct evidence of the competition between nematic and superconducting phases. Above the nematic phase transition, the orthorhombic lattice distortion induced by uniaxial stress plays the role of the conjugate field of the primary nematic order parameter. One application of this idea is the measurement of the bare nematic susceptibility, in which the induced electronic anisotropy is measured above the phase transition under a constant anisotropic strain (*17–22*). The divergence of this bare nematic susceptibility demonstrates unambiguously that the structural transition is electronically driven (*23*). Significantly, a diverging nematic susceptibility is also observed in a wide range of optimally-doped iron-based superconductors, suggesting the correlation of optimal $T_c$ with nematic quantum critical fluctuations (*24*).

In the present study, we show that anisotropic strain is indeed an effective knob to tune the superconducting transition in $Ba(Fe_{1-x}Co_x)_2As_2$. We find that the $T_c$ of underdoped and near-optimally doped regions can be strongly suppressed by both compressive and tensile uniaxial stress. In the underdoped case this suppression is sufficient to induce a superconductor to metal quantum phase transition with less than one percent strain. Intriguingly, the sensitivity of $T_c$ to strain vanishes rapidly in the overdoped side of the phase diagram. This strong doping dependence of strain sensitivity may provide a quantitative test of the superconducting pairing mechanism.

Resistivity measurements were performed as a function of temperature and applied uniaxial stress, with current flowing parallel to the direction of the applied stress along the Fe-Fe bonding direction. In this configuration the uniaxial stress induces a $B_{2g}$ anisotropic strain that couples linearly to nematicity. We used a strain cell (Fig. S1) capable of applying large, tunable uniaxial stress at low temperatures (*25*). The uniaxial stress is controlled by the displacement of two strain cell plates across which the sample is mounted. The relative change of the size of gap between the plates, $\varepsilon_{disp}$, was measured by resistive strain gauges, and corresponds to a tensile (compressive) uniaxial stress for positive (negative) values. Nevertheless, $\varepsilon_{disp}$ is distinct from the actual lattice distortion due to the imperfect strain transmission and the formation of twin domains in the underdoped samples. The actual lattice distortion was either measured directly by x-ray diffraction or calculated by finite element simulations (*26*).

We focus first on the *x* = 0.042 composition which lies in the underdoped region of the phase diagram (Fig. 1A). With decreasing temperature, the free-standing sample undergoes nematic (structural), magnetic, and superconducting phase transitions at $T$ = 77, 69 and 13K respectively (Fig. 1B). Thus, the application of uniaxial stress is expected to enhance an orthorhombicity that is already present at zero stress due to the nematic transition. The resistivity as a function of temperature under different applied stress is shown in Fig. 1C-F. Inspection of this data shows that for small $\varepsilon_{disp}$, the value of the resistivity above the superconducting transition changes rapidly while the superconducting transition temperature changes little. In this range of $\varepsilon_{disp}$ the only effect of uniaxial stress is to align nematic domains without changing the lattice constants, and the strong modulation of resistivity is due to the large resistance anisotropy associated with the orthorhombic phase (*27*). Then for $\varepsilon_{disp} > 2.7 \times 10^{-3}$ and $\varepsilon_{disp} < -0.9 \times 10^{-3}$, at which point the sample is fully detwinned and the crystal lattice is further distorted by the uniaxial stress, the superconducting transition is dramatically suppressed for both compressive (Fig. 1C) and tensile (Fig. 1D) stress. The effect of the stress on the superconducting transition is even more striking when the resistivity is shown on a logarithmic scale (Fig. 1E-F). At the highest values of $\varepsilon_{disp}$, we measure a nonzero resistivity all the way down to the base temperature of our system.

The broad resistive transition at large values of $\varepsilon_{disp}$ raises the question of how to define the phase boundary of the superconducting state. Due to the extreme sensitivity of $T_c$ to $\varepsilon_{disp}$ , even small strain gradients or slight strain drift as the temperature is changing may significantly broaden the transition. To mitigate these effects, we investigated the behavior of the $x = 0.042$ composition under conditions of fixed temperature and variable $\varepsilon_{disp}$. In addition, in order to confirm that the superconductivity in the bulk of the sample has been suppressed, we also measured the AC magnetic susceptibility as a function of $\varepsilon_{disp}$ (Fig. 2A-B), which is a technique already shown to be an effective method for measuring superconducting transitions of strained samples (*28, 29*). At high temperatures, there is near zero signal for all $\varepsilon_{disp}$, indicating the sample is in the normal state. Then at low temperatures and low $\varepsilon_{disp}$, there is a diamagnetic signal consistent with a superconducting Meissner screening state. By applying either tensile or compressive stress, this diamagnetic signal is suppressed, and there is a sharp peak in the imaginary part of the signal which we take to be clear evidence of a thermodynamic transition out of the superconducting state. This behavior mirrors what is seen in the corresponding resistivity data at fixed temperature and variable $\varepsilon_{disp}$, which is shown in Fig. 3C. *I-V* curves performed at fixed temperature and $\varepsilon_{disp}$ (Fig. S2) reveal suppressed critical currents and nonlinear behavior near the zero resistivity ($\rho = 0$) phase boundary, and that nonlinearity evolves into linear Ohmic behavior with increasing $\varepsilon_{disp}$, confirming the full recovery of a metallic state.

To further elucidate how the superconducting state responds to the lattice distortion induced by $\varepsilon_{disp}$, we performed x-ray diffraction measurements at beamline 6-ID-B at the Advanced Photon Source (APS) where we have a sample environment that can simultaneously measure electrical transport and apply uniaxial stress while performing the x-ray diffraction measurements (*26*). For a finite range of $\varepsilon_{disp}$ around zero, both the in-plane and out-of-plane lattice constants do not change as the crystal is being detwinned (Fig. S4A,C,D) as the displacement is being absorbed by shifting domain populations and the orthorhombicity is flat. For higher $\varepsilon_{disp}$, the lattice constants begin to change, and the orthorhombicity ($\delta = \frac{a-b}{a+b}$) increases with both compressive and tensile stress. Eventually, at a critical orthorhombicity a finite resistance state is recovered (Fig. 2D). In addition to direct visualization of the evolution of lattice constants, this measurement also demonstrates that the strain of the sample is homogenous by monitoring the sharpness of the Bragg peaks (Fig. S4B). Overall, the compilation of transport and magnetic susceptibility data for this composition (Fig. 2E) clearly demonstrates a superconductor to metal quantum phase transition (QPT) as a function of $\varepsilon_{disp}$.

The slightly underdoped composition $x = 0.06$ exhibits similar behavior as the $x = 0.042$ composition (Fig. S5). The question arises of whether a natural structural distortion or static magnetic order are necessary ingredients to observe this rapid suppression of $T_c$. To answer this question, we performed the same resistivity measurements under uniaxial stress on an optimally doped ($x = 0.071$) sample. For tetragonal samples, the relevant crystal point group is $D_{4h}$, and the type of strain induced by uniaxial stress can be decomposed as the sum of two irreducible representations $\varepsilon_{A1g}$, corresponding to non-symmetry breaking strains such as volume expansion and change of tetragonality, and $\varepsilon_{B2g} = \frac{1}{2}(\varepsilon_{xx} - \varepsilon_{yy})$ which breaks the fourfold rotational symmetry. By symmetry considerations, to lowest order the superconducting $T_c$ can only depend quadratically on $\varepsilon_{B2g}$ but can depend linearly on $\varepsilon_{A1g}$. Mathematically, we have

$$T_c(\varepsilon) = T_c^0(1 + \beta\varepsilon_{A1g} - \alpha\epsilon_{B2g}^2)$$

where $\alpha$ and $\beta$ parametrize the dimensionless sensitivity of $T_c$ to $\varepsilon_{A1g}$ and $\varepsilon_{B2g}$ respectively. Consequently, one would expect a monotonic dependence of $\varepsilon_{disp}$ if the $T_c$ is primarily determined by $\varepsilon_{A1g}$, and a symmetric response to both positive and negative $\varepsilon_{disp}$ if the effect of $\varepsilon_{B2g}$ dominates. Remarkably, in the optimally doped sample we observe a strong suppression of superconductivity for both positive and negative $\varepsilon_{disp}$ – a nearly five-fold reduction of $T_c$ at about 1% $\varepsilon_{disp}$. Although the effect of strain on superconductivity was inferred in several previous works (*30*, *31*), here we unambiguously demonstrate the dominant effect of $\varepsilon_{B2g}$ and hence confirm the dual role played by the broken rotational symmetry phase.

This extreme sensitivity of the superconducting state to $B_{2g}$ strain is truncated past optimal doping (Fig. 3A); for a slightly overdoped sample with a similar transition temperature ($x = 0.088$), 0.5% of $\varepsilon_{disp}$ induced by compressive stress only reduces $T_c$ by 12%, compared with a 50% reduction in the optimally doped sample for the same $\varepsilon_{disp}$. Further into the overdoped regime, ($x = 0.113$) the response of $T_c$ to $\varepsilon_{disp}$ is even smaller in magnitude and no longer symmetric for tensile and compressive stress. By calculating the amount of $\varepsilon_{B2g}$ at a given $\varepsilon_{disp}$ using linear elasticity theory (*26*) and recent systematic measurements of the elastic constants (*32*), we plot $T_c$ against the purely antisymmetric $\varepsilon_{B2g}$ in Fig. 3B. It is clear from the behavior of $T_c$ that there is a crossover from a $\varepsilon_{B2g}$-dominated response to a $\varepsilon_{A1g}$-dominated response of the superconducting state as the doping level is increased.

The drastically different behavior between underdoped, optimally doped, and overdoped samples is rather intriguing. In the simplest view, the suppression of the $T_c$ should be directly related to the amount of static nematic order induced by a fixed amount of $\varepsilon_{B2g}$, which is measured by the nematic susceptibility. In Fig. 3C we plot the doping dependence of the elastoresistivity coefficient *-2m₆₆* measured just above $T_c$, which is proportional to the nematic susceptibility, and the strain sensitivity of $T_c$, characterized by the coefficient $\alpha$ as defined above. The coefficient $\alpha$ shows a much stronger doping dependence compared to $2m_{66}$. Motivated by a recent theoretical work on nematic-mediated superconductors (*33*), we also extracted the power law relations between these two quantities. From a limited number of data points, the two quantities show a perfect power law dependence with an exponent of 1.9 (Fig. S7). We also note that static and/or fluctuating antiferromagnetic order may play a non-trivial role in the determination of $\alpha$, as previous neutron scattering experiments revealed the increase of $T_N$ and enhanced magnetic moments under uniaxial pressure in the underdoped iron pnictides (*31, 34*). A systematic study of the strain dependence of spin fluctuations in the overdoped and optimally doped compounds will further elucidate the contribution to superconducting pairing from different degrees of freedom.

In a broader view, regardless of the exact mechanism at play here, these measurements reveal an unprecedented tunability of $T_c$ by lattice deformation in a bulk superconductor. To put iron pnictides into context, we list the $dT_c/d\varepsilon$ of several common superconductors in Table S3 (*29, 35–38*). The $dT_c/d\varepsilon$ of the optimally doped sample is much larger than any other known superconductor. Such tunability allows us to construct the doping-strain-temperature ($x$-$\varepsilon$-$T$) phase diagram which we show schematically in Fig. 3D. For a broad range of doping near the composition-tuned nematic quantum critical point, the superconducting state is extremely sensitive to $B_{2g}$ strain which acts to close the superconducting dome in the $\varepsilon$-$T$ plane, generating a line of superconductor-metal QPTs in the zero-temperature limit.

The superconductor to metal (or insulator) transition is one of the most studied quantum phase transitions in condensed matter physics (*39–41*). Nevertheless, previous experimental studies mostly restricted to two-dimensional system, possibly due to the ease of continuous tuning $T_c$ in thin films or exfoliated thin flakes. Our work presents a new platform to study the superconductor-metal transition in a three-dimensional crystal with an in-situ tunable strain. The three-dimensional sample may allow experimental probes that were previously inaccessible in a two-dimensional system, such as thermodynamic measurement, which may shed new light onto the recently proposed "anomalous metal" state (*42*).

**Supplementary Materials:**

Materials and Methods

Supplementary Text

Supplementary Figures S1-S7

Tables S1-S3

**Acknowledgments:** We thank Xiaodong Xu, David Cobden, Boris Spivak, Steve Kivelson, and Cenke Xu for valuable discussion. **Funding:** This work was mainly supported by NSF MRSEC at UW (DMR-1719797) and the Gordon and Betty Moore Foundation's EPiQS Initiative, Grant GBMF6759 to J.-H.C. The development of strain instrumentation is supported as part of Programmable Quantum Materials, an Energy Frontier Research Center funded by the U.S. Department of Energy (DOE), Office of Science, Basic Energy Sciences (BES), under award DE-SC0019443. The integration of x-ray diffraction with *in-situ* strain is supported by the Air Force Office of Scientific Research Young Investigator Program under Grant FA9550-17-1-0217. J.H.C. acknowledges the support of the David and Lucile Packard Foundation and the State of Washington funded Clean Energy Institute. **Author contributions:** P.M., J.S., J.M., Q.J., and Z.L. grew the samples. P.M., Q.J., and J.S. did the experiments. P.R., J.-W.K., and J.L. helped conceive and design the XRD measurements at the APS. P.W. and P. M. performed the finite element analysis. P.M. analyzed the data. J.H.C. supervised the project. All authors contributed extensively to the interpretation of the data and the writing of the manuscript.

**Competing interests:** Authors declare no competing interests. **Data and materials availability:** All data is available in the main text or the supplementary materials or upon reasonable request.

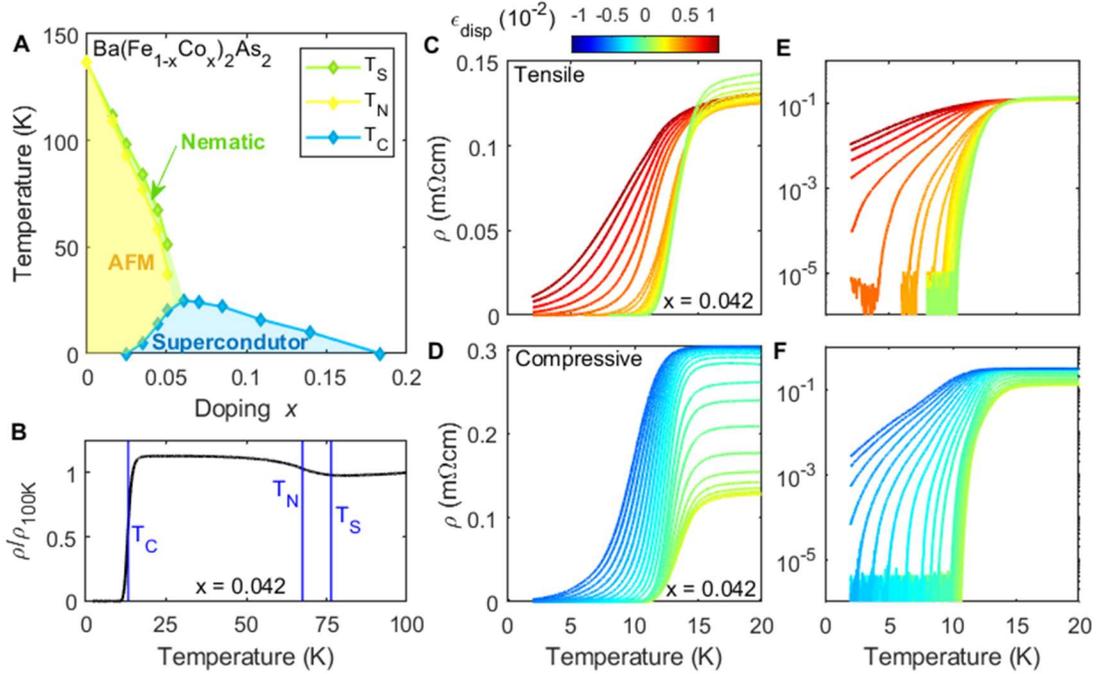

**Fig. 1. Background and superconducting transition in Ba(Fe$_{0.958}$Co$_{0.042}$)$_2$As$_2$ under uniaxial stress (A)** Temperature-composition phase diagram (*14*) of Ba(Fe$_{1-x}$Co$_x$)$_2$As$_2$ **(B)** Resistivity as a function of temperature for the $x$ = 0.042 composition. Blue vertical lines indicate the locations of the various phase transitions. **(C,D)** Resistive signature of the superconducting transition in the $x$ = 0.042 composition as a function of temperature under different amounts of applied uniaxial compressive (C) and tensile (D) stress **(D,F)** Same as in (C+D) with resistivity on a log scale

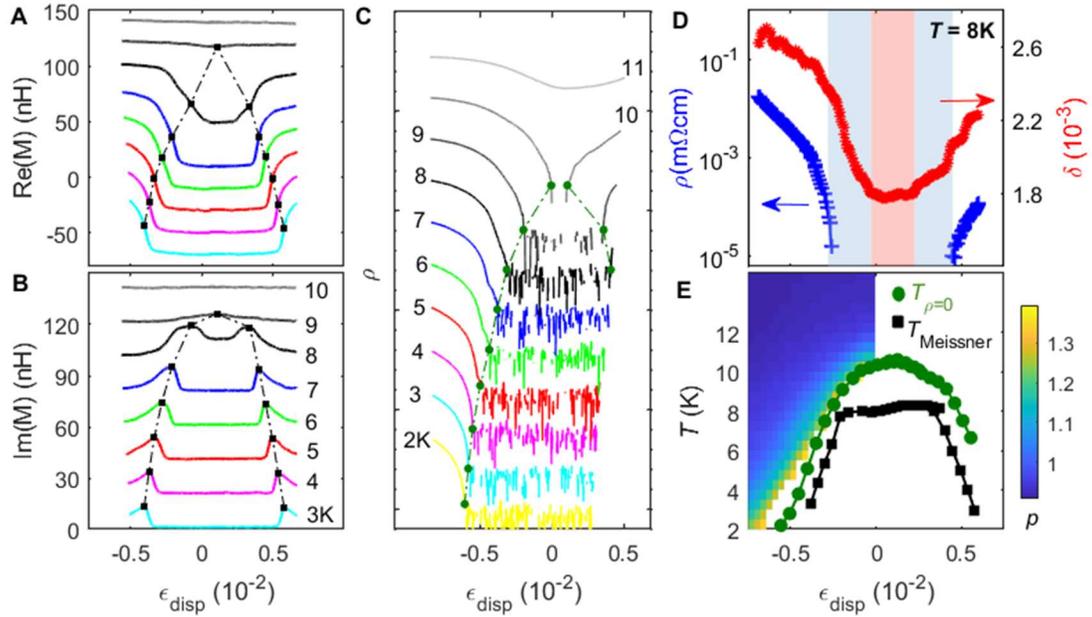

**Fig. 2. Strain-tuned superconductor to metal transition in Ba(Fe$_{0.958}$Co$_{0.042}$)$_2$As$_2$**

Real **(A)** and imaginary **(B)** parts of the mutual inductance $M$ of the susceptometer coils at fixed temperatures. Black squares indicate the strain at which the Meissner effect is suppressed, and curves are offset for clarity **(C)** Resistivity on logarithmic scale under strain at fixed temperatures. Green circles indicate the strain at which the resistivity is non-zero and curves at different temperatures are offset for clarity **(D)** Comparison of the resistivity (blue, log scale) and orthorhombicity (red) during the SC-metal transition **(E)** Strain-temperature phase diagram for the $x = 0.042$ composition. The color bar indicates the power $p$ of the $I$-$V$ characteristic as described in the text.

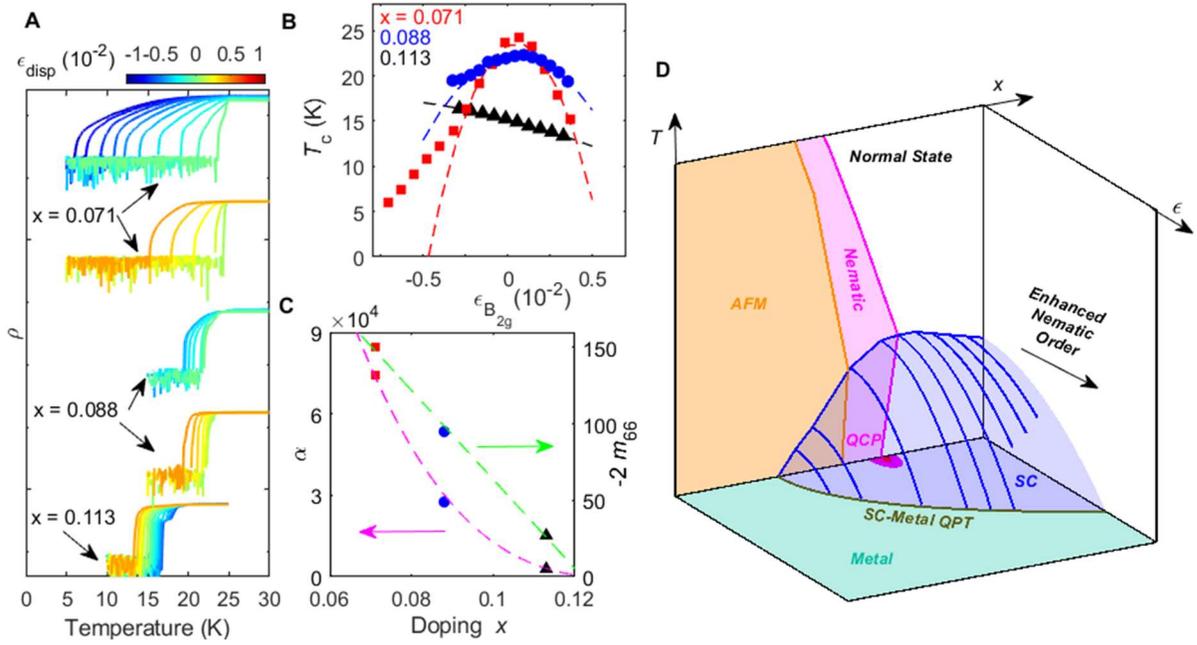

**Fig. 3. Strained superconducting transition of optimally and over doped Ba(Fe$_{1-x}$Co$_x$)$_2$As$_2$ and schematic doping-strain-temperature ($x$-$\epsilon$-$T$) phase diagram (A)** Resistivity as a function of temperature under uniaxial stress for optimally and overdoped compositions. Resistivity is on a log scale and sets of curves that overlap are offset for clarity **(B)** Extracted $T_c$ as a function of $\epsilon_{B_{2g}}$ strain. Dotted lines are second order polynomial fits to the low strain data (fit parameters are tabulated in Table S2). **(C)** Doping dependence of the (right) nematic susceptibility $-2m_{66}$ at $T =$ 30K and (left) normalized quadratic coefficient $\alpha = \frac{-d^2T_c}{d\varepsilon^2}\frac{1}{T_c}$. Green and magenta dotted lines are guides to the eye. **(D)** Schematic doping-strain-temperature ($x$-$\epsilon$-$T$) phase diagram. The magenta region indicates the electronic nematic phase which breaks the $C_4$ rotational symmetry, ending in a quantum critical point (QCP) inside the superconducting dome (blue region). At fixed doping, anisotropic strain that enhances the broken symmetry suppresses the superconducting $T_c$ (solid blue lines), generating another superconducting dome tuned by strain instead of doping. The closing of this dome gives way to a metallic phase (green region) and generates a line of strain-tuned superconductor-to-metal quantum phase transitions (QPTs) in the $x$-$\epsilon$ plane (dark green line).

# Supplementary Materials for

## Drastic suppression of superconducting $T_c$ by anisotropic strain near a nematic quantum critical point

Paul Malinowski[1], Qianni Jiang[1], Joshua Sanchez[1], Joshua Mutch[1], Zhaoyu Liu[1], Preston Went[1], Jian Liu[2], Philip Ryan[3,4], Jong-Woo Kim[3], Jiun-Haw Chu[1*].

Correspondence to: jhchu@uw.edu (JHC)

**Affiliations:**

[1]Department of Physics, University of Washington, Seattle, Washington 98195, USA.

[2]Department of Physics, University of Tennessee, Knoxville, Tennessee 37996, USA.

[3]Advanced Photon Source, Argonne National Laboratories, Lemont, Illinois 60439, USA.

[4]School of Physical Sciences, Dublin City University, Dublin 9, Ireland.

*Correspondence to: jhchu@uw.edu (J.-H.C)

**This PDF file includes:**

Materials and Methods

Supplementary Text

Figs. S1 to S7

Tables S1 to S3

**Materials and Methods**

<u>Sample preparation</u>

Single crystals of $Ba(Fe_{1-x}Co_x)_2As_2$ were grown out of FeAs flux as described previously (*15-16*). The crystals preferentially break along the $(100)_{Tet}$ and $(010)_{Tet}$ directions when cleaved which allowed for determination of the crystallographic directions ("Tet" subscript signifies that the Miller indices are referenced from the high-temperature tetragonal lattice). Composition of the samples was determined using energy dispersive spectroscopy (EDS). The crystals were cleaved into thin bars with the $(110)_{Tet}$ direction being the longest dimension. Electrical contacts were made in a typical 4-point configuration with sputtered gold pads and silver epoxy.

<u>Strain device and strain determination</u>

For applying uniaxial stress, a home-built three-piezo device was used (Fig. S1). In this type of strain cell (*25*), a crystal is glued so that a portion of its length is suspended across a gap in a titanium scaffolding which has three piezo stacks attached in such a way that extension (compression) of the outer stacks and compression (extension) of the inner stack will cause the crystal to experience tensile (compressive) strain. In this configuration, thermal expansion of the piezo stacks does not affect the sample because of their symmetric arrangement, and the large ratio of the length of the gap over the length of the piezo stacks allows for large strains of 1% or more to be applied to the sample even at cryogenic temperatures.

Even though the thermal expansion of the piezo stacks is eliminated in this configuration, knowing the displacement of the mounting plates is not enough to know the $\epsilon_{disp}$ experienced by the sample because of two facts:

1. The presence of differential thermal contraction between the sample and the titanium from which the plates are made. This requires a separate determination of the zero-strain point on the sample.

2. A non-perfect strain transmission though the epoxy used to affix the sample to the mounting plates

To address the first point and determine the point of zero strain for the sample, we utilize the fact that the $Ba(Fe_{1-x}Co_x)_2As_2$ system demonstrates relatively large gauge factors, meaning that the resistivity is very sensitive to strain. Thus, to determine zero strain, we would first measure the temperature dependence of the resistivity of a crystal in the free-standing state, not mounted on the strain cell. Then, once mounted on the strain cell, we would sweep the voltage on the stacks until the resistance matched the resistance of the free-standing state.

To address the second point, we performed finite element analysis using the *ANSYS Academic Research Mechanical 19.1* software to model the strain transmission utilizing the elastic properties of both the sample, which were taken from reference (32), and the elastic properties of the mounting epoxy (Loctite Stycast 2850FT) which is available from the manufacturer. The result of these calculations is a coefficient $\alpha$ which indicates the percentage of the strain transmitted to the sample. Typical values of $\alpha$ are 0.8-0.9.

Once the zero-point and strain transmission factors are known, monitoring of the strain experienced by the sample was done using a resistive strain gauge. For the resistive strain gauge, the gauge (SS-150-124-15P, Micron Instruments, Simi Valley, CA) is glued onto the back of the center piezo stack. These gauges have a known room temperature gauge factor provided by the manufacturer, where the gauge factor $g$ is defined through the equation

$$\Delta\rho / \rho = \frac{\rho(\epsilon) - \rho(\epsilon = 0)}{\rho(\epsilon = 0)} = g * \epsilon_{piezo}$$

The temperature dependence of the gauge factor was calibrated ourselves prior to the measurements. For this type of strain gauge, $g = 80$ at room temperature and $g = 165$ at our base temperature of 2K and follows a linear temperature dependence for intermediate temperatures. Here $\epsilon_{piezo} = \frac{\Delta L}{L_{piezo}}$ represents the strain experienced by the piezo itself. Because we always drive the outer piezo stacks exactly opposite to the inner piezo stack, the change of length of the gap across which the sample is suspended is equal to twice the change of length of a single piezo, so we have $\epsilon_{disp} = \alpha \frac{2\Delta L}{L_{gap}} = \alpha \frac{2*L_{piezo}}{L_{gap}} \epsilon_{piezo}$, where $\alpha$ is the strain transmission factor discussed above. The quantity $\frac{2*L_{piezo}}{L_{gap}}$ is the mechanical advantage that allows for large strains to be applied to the sample, and is equal to 36 for typical values of $L_{piezo} = 9mm$ and $L_{gap} = 0.5mm$. Combining the above equations gives us that the strain experienced by the sample in terms of the strain gauge resistance $R^{SG}$, which is the quantity that we directly measure, is

$$\epsilon_{disp} = \alpha * \frac{2 * L_{piezo}}{L_{gap}} * \frac{R^{SG} - R_0^{SG}}{R_0^{SG}}$$

where $R_0^{SG}$ is the value of the strain gauge resistance at the point where the crystal is in the zero-strain state as determined above.

Magnetic susceptibility

We follow procedures similar to those used in (*28-29*). An AC current at frequencies ~100 Hz is sent through the excitation coil (~100 turns) which has a diameter of approximately 2 mm and is placed approximately 1 mm above the sample and generates a field ~1 Oe at the sample. The signal from a second, smaller coil (~100 turns) placed directly above the strained region of the sample is fed into a SR554 transformer preamplifier and then into a SR830 lock-in amplifier listening to the frequency of the excitation current. Both the in-phase and out-of-phase components are then monitored.  The AC susceptibility is in general a complex quantity, $\chi = \chi' - i\chi''$.  Assuming an excitation field of the form $H = H_0 cos(2\pi f t)$, then the real part $\chi'$ represents the reversible magnetization that will oscillate in phase with the excitation field, while the imaginary part $\chi''$ represents any irreversible magnetization processes.

Extraction of the exact numerical values for $\chi'$ and $\chi''$ would require detailed modeling and solution of Maxwell's equations for the exact geometry, but the exact values are not important for our conclusions. Furthermore, the geometric arrangement at hand leads to a large background signal that arises solely from the inductive coupling of the excitation and secondary coils and is unrelated to the magnetic response of the sample.  Therefore, we measured the mutual inductance of the coils mounted on the cell both with and without the sample, and then subtracted the two in order to isolate the response from just the sample. We then plot the measured "effective" mutual inductance of the set of coils, which is equal to

$$M = \frac{\Phi_{sample}}{I} = \frac{\Phi_{total} - \Phi_{background}}{I} = \frac{\omega(V_{total} - V_{background})}{I}$$

where $\omega$ and $I$ are the measurement frequency and excitation current respectively, and $V_{total}$ and $V_{background}$ are the signals measured on the secondary coil.

X-ray diffraction

High energy x-ray diffraction measurements were performed at beamline 6-ID-B at the Advanced Photon Source with an energy of 11.215 keV and wavelength of 1.10552 Å.  The strain cell was mounted on the cold finger of a closed cycle cryostat allowing for temperature control between 7 and 300K.  Electrical contacts made on the underside of the sample allowed for simultaneous resistivity and diffraction measurements under uniaxial stress without blocking the path of the x-rays.  Below the structural/nematic transition, the presence of orthorhombic twin domains causes splitting of the Bragg peaks sensitive to the in-plane lattice constants along the Fe-Fe bonds and uniaxial stress acts to (Fig. S4A). By measuring the $(2\ 2\ 12)_{Tet}$, $(1\ \text{-}1\ 14)_{Tet}$, and $(0\ 0\ 14)_{Tet}$ Bragg peaks, we were able to extract the orthorhombic $a,b$ in-plane lattice constants both parallel and perpendicular to the direction of applied stress, as well as the out of plane $c$ lattice constant (Fig. S4B,D). Measurement of the in-plane lattice constant allows for determination of the orthorhombicity $\delta = \frac{a-b}{a+b}$ as a function of applied stress as shown in the main text. In addition, the x-ray diffraction measurements under stress confirm that the crystal is experiencing a state of homogenous strain; the widths of the Bragg peaks under both compressive and tensile strain are comparable to the width of the Bragg peak near zero stress (Fig. S4B). This, coupled with the fact that we are using a large beam spot (~250 μm) that covers the whole width of the sample, attests to homogenous strains.

<u>Definition and extraction of T$_c$</u>

Throughout the text, the superconducting $T_c$ is defined as the point where the resistivity of the sample is equal to zero. In the superconducting state, a resistivity measurement made with a lock-in amplifier will yield a signal that consists of a base noise level that is zero on average. One consistent way of defining the point of "zero resistance" is to find the root mean square (RMS) of the noise far below the superconducting transition, and then define the transition to zero resistance to be the first point at which the signal falls below that level (Fig. S3). This is the definition that we use throughout this work.

<u>Relation between applied strain and purely anisotropic strain</u>

The quantity directly measured in our setup, $\epsilon_{disp}$, is the strain along the direction of the current flow, and is not the same as the quantity $\epsilon_{B_{2g}} = \frac{1}{2}(\epsilon_{aa} - \epsilon_{bb})$ that is of the same symmetry as the natural lattice distortion and consequently is the conjugate field that directly couples to the nematic order parameter. Assuming we have a tetragonal lattice and letting $a$ and $b$ represent the lengths parallel and perpendicular respectively to the direction of strain, then we have $\epsilon_{bb} = -\nu_{ab}\epsilon_{aa}$ where $\nu_{ab}$ is the appropriate Poisson ratio.  This then gives

$$\epsilon_{B_{2g}} = \frac{1}{2}(\epsilon_{\text{aa}} - \epsilon_{\text{bb}}) = \frac{1}{2}\epsilon_{aa}(1 + \nu_{ab})$$

The quantity $\nu_{ab}$ can be calculated from the coefficients of the compliance tensor. Reference *32* has detailed temperature and doping dependence of the stiffness tensor of the Ba(Fe$_{1-x}$Co$_x$)$_2$As$_2$ system extracted from ultrasound measurements.  However, those coefficients are in the basis of the tetragonal lattice.   For a tetragonal crystal, the stiffness tensor takes the form

$$C = \begin{pmatrix} C_{11} & C_{12} & C_{13} & 0 & 0 & 0 \\ C_{12} & C_{11} & C_{13} & 0 & 0 & 0 \\ C_{13} & C_{13} & C_{33} & 0 & 0 & 0 \\ 0 & 0 & 0 & C_{44} & 0 & 0 \\ 0 & 0 & 0 & 0 & C_{44} & 0 \\ 0 & 0 & 0 & 0 & 0 & C_{66} \end{pmatrix}$$

The compliance tensor is the inverse of the stiffness tensor, so we have

$$S = C^{-1} = \begin{pmatrix} \dfrac{-C_{13}{}^2 + C_{11}C_{33}}{A} & \dfrac{-C_{13}{}^2 - C_{12}C_{33}}{A} & \dfrac{C_{13}}{B^+} & 0 & 0 & 0 \\[2mm] \dfrac{-C_{13}{}^2 - C_{12}C_{33}}{A} & \dfrac{-C_{13}{}^2 + C_{11}C_{33}}{A} & \dfrac{C_{13}}{B^+} & 0 & 0 & 0 \\[2mm] \dfrac{C_{13}}{B^+} & \dfrac{C_{13}}{B^+} & \dfrac{C_{13}}{B^-} & 0 & 0 & 0 \\[2mm] 0 & 0 & 0 & \dfrac{1}{C_{44}} & 0 & 0 \\[2mm] 0 & 0 & 0 & 0 & \dfrac{1}{C_{44}} & 0 \\[2mm] 0 & 0 & 0 & 0 & 0 & \dfrac{1}{C_{66}} \end{pmatrix}$$

where $A = \dfrac{-C_{13}{}^2 + C_{11}C_{33}}{(C_{11} - C_{12})(-2C_{13}{}^2 + C_{33}(C_{11} + C_{12}))}$ and $B^\pm = \pm(2C_{13}{}^2 - C_{33}(C_{11} + C_{12}))$.

This matrix is still in the basis of the tetragonal lattice, whereas we need the matrix in the basis where the in-plane axes are rotated 45 degrees with respect to this basis. We then have

$S' = K^{-T} S K^{-1}$, where $K$ is the rotation matrix given by (33)

$$K = \begin{pmatrix} c^2 & c^2 & 0 & 0 & 0 & 2cs \\ c^2 & c^2 & 0 & 0 & 0 & -2cs \\ 0 & 0 & 1 & 0 & 0 & 0 \\ 0 & 0 & 0 & c & s & 0 \\ 0 & 0 & 0 & -s & c & 0 \\ -cs & cs & 0 & 0 & 0 & c^2 - s^2 \end{pmatrix}$$

and $c = \cos(45°), s = \sin(45°)$.

This then gives the rotated compliance tensor:

$$S' = \begin{pmatrix} \dfrac{C_{33}}{-2B^-} + \dfrac{1}{4C_{66}} & \dfrac{C_{33}}{-2B^-} - \dfrac{1}{4C_{66}} & \dfrac{C_{13}}{B^+} & 0 & 0 & 0 \\ \dfrac{C_{33}}{-2B^-} - \dfrac{1}{4C_{66}} & \dfrac{C_{33}}{-2B^-} + \dfrac{1}{4C_{66}} & \dfrac{C_{13}}{B^+} & 0 & 0 & 0 \\ \dfrac{C_{13}}{B^+} & \dfrac{C_{13}}{B^+} & \dfrac{C_{11+}C_{12}}{B^-} & 0 & 0 & 0 \\ 0 & 0 & 0 & \dfrac{1}{2C_{44}} & 0 & 0 \\ 0 & 0 & 0 & 0 & \dfrac{1}{2C_{44}} & 0 \\ 0 & 0 & 0 & 0 & 0 & \dfrac{2}{C_{11} - C_{12}} \end{pmatrix}$$

Finally, the Poisson ratio of interest $\nu_{ab} = -\dfrac{\epsilon_{bb}}{\epsilon_{aa}}$ is then equal to

$$\nu_{ab} = -\frac{\epsilon_{bb}}{\epsilon_{aa}} = \frac{-S_{12}}{S_{11}} = \frac{-C_{13}^2 + \frac{1}{2}(C_{11} + C_{12})C_{33} - C_{33}C_{66}}{-C_{13}^2 + \frac{1}{2}(C_{11} + C_{12})C_{33} + C_{33}C_{66}}$$

Table S1 lists the elastic constants for the optimally and overdoped samples along with the calculated Poisson ratio.

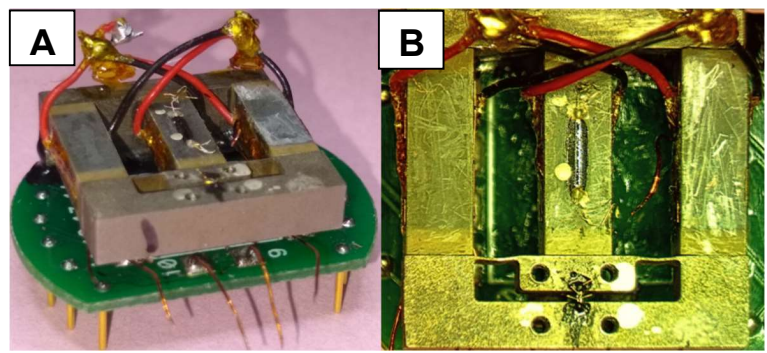

**Fig. S1. Device for applying uniaxial stress (A)** Side view **(B)** Top view

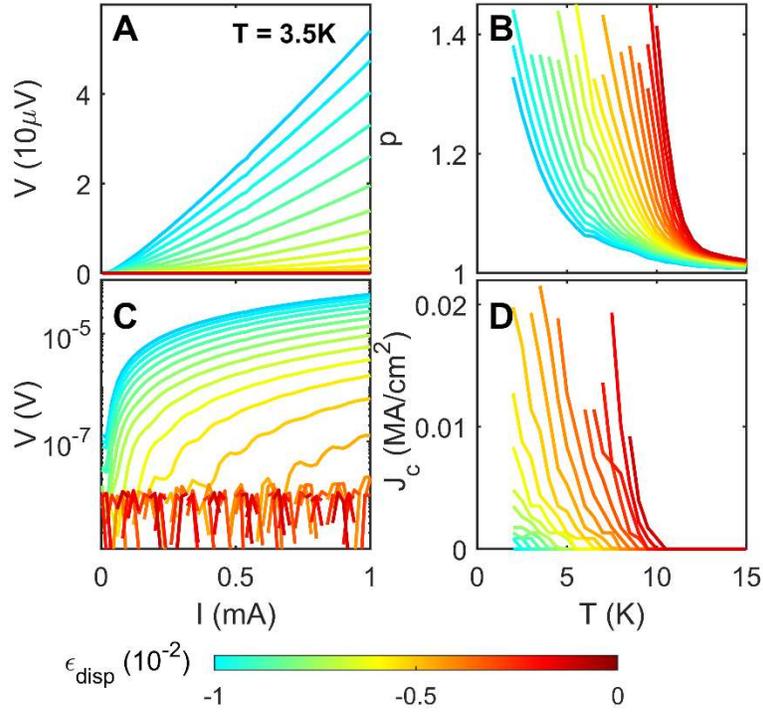

**Fig. S2. Nonlinear IV characteristics and critical currents as a function of strain for the *x* = 0.042 composition (A)** Characteristic set of *I-V* curves showing nonlinear behavior near the $\rho = 0$ phase boundary. **(B)** Temperature and power dependence of the power *p* defined as $V = I^p$. **(C)** Same as (A) with voltage on a log scale demonstrating suppression of the critical current with increasing strain. **(D)** Temperature and strain dependence of the superconducting critical current density.

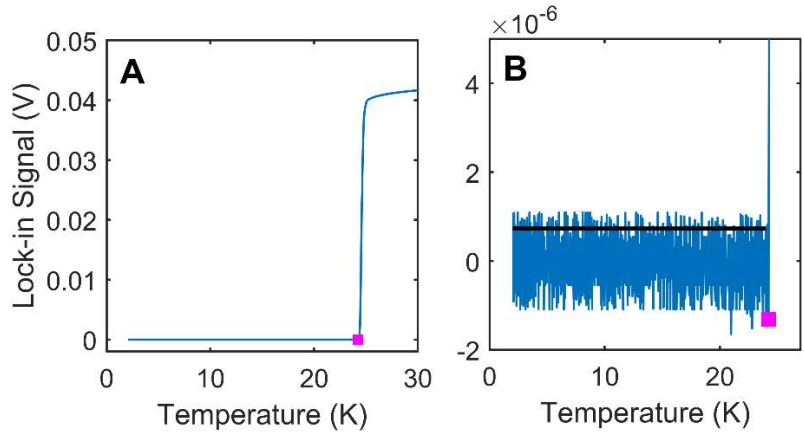

**Fig. S3 Definition of $T_c$ as point of zero resistance (A)** Typical resistive signature of the superconducting transition for the $x = 0.071$ composition with the point of zero resistance marked in magenta **(B)** Same as the data in the left panel but zoomed in close to the transition. The black line indicates the root mean square of the noise level as discussed in the supplementary text.

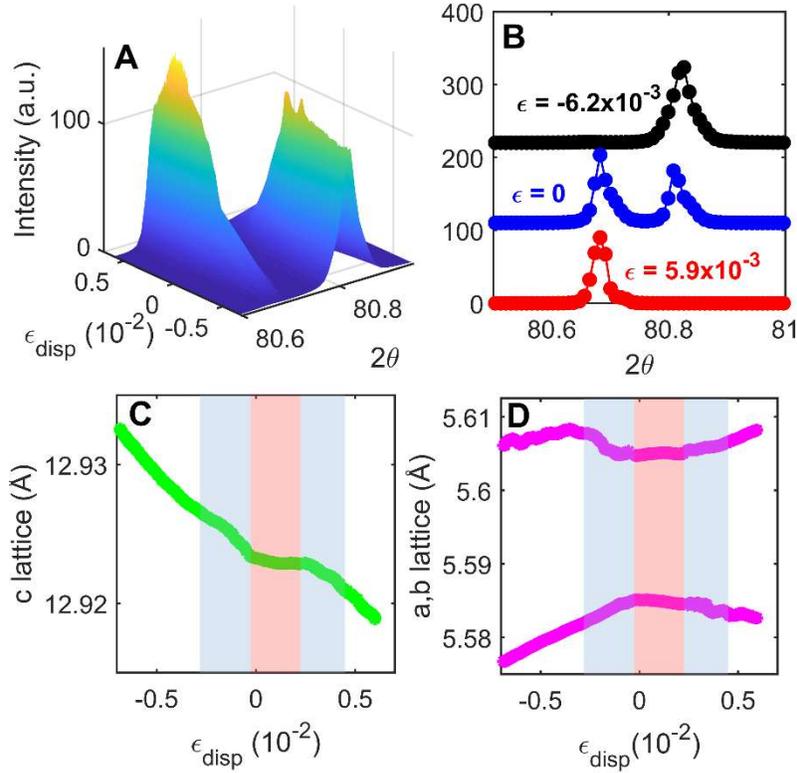

**Fig. S4. X-ray diffraction measurements of the *x* = 0.042 under uniaxial stress (A)** Splitting of the (2 2 12) Bragg peak showing the two structural domains and control of the domain population with applied stress. **(B)** $2\theta$ scan of the (2 2 12) peak under compressive (black), zero (blue), and tensile (red) stress. **(C)** *c*-lattice constant and **(D)** in-plane lattice constants as a function of strain. Red shaded region indicates the region where the crystal is being detwinned and the lattice constants are relatively constant. Blue shaded region indicates the additional macroscopic strain needed to recover a finite resistance state.

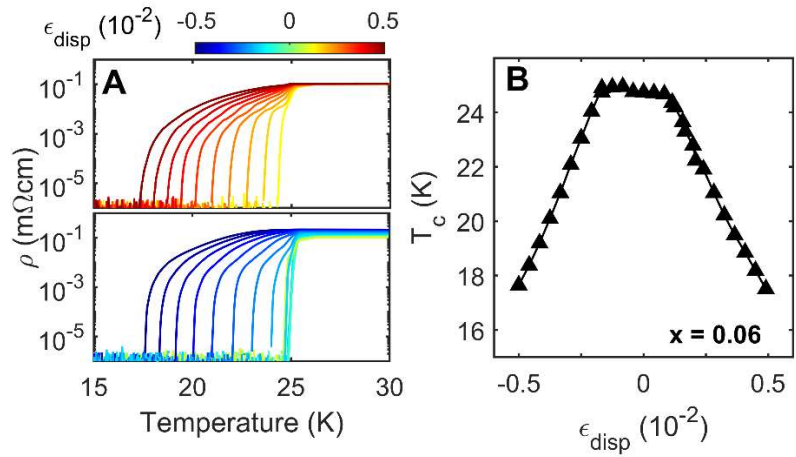

**Fig. S5 Effect of uniaxial stress on the superconducting transition for the *x* = 0.06 composition (A)** Resistivity as a function of temperature under different amounts of compressive and tensile strain **(B)** Extracted superconducting $T_c$ as a function of applied strain.

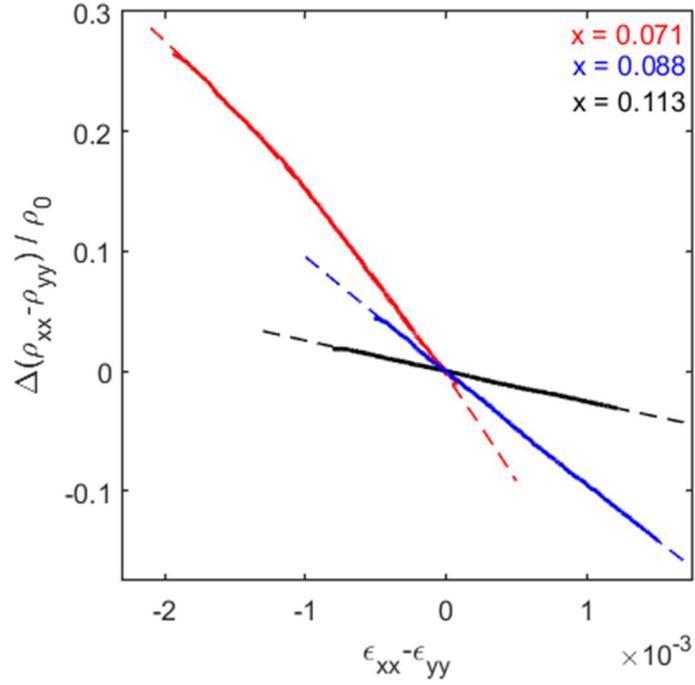

**Fig. S6 Measurement of doping dependence of nematic susceptibility just above $T_c$.**
Measurement of the anisotropic resistive response $\frac{\Delta(\rho_{xx} - \rho_{yy})}{\rho_0}$ as a function of the anisotropic strain $\epsilon_{xx} - \epsilon_{yy}$ at $T = 29\mathrm{K}$. The two components $\rho_{xx}$ and $\rho_{yy}$ of the resistivity tensor are measured on the same sample as a function of strain using the modified Montgomery method with the sample glued on the sidewall of a single piezo-stack as described in (*24*). The dotted lines are quadratic fits to the data.

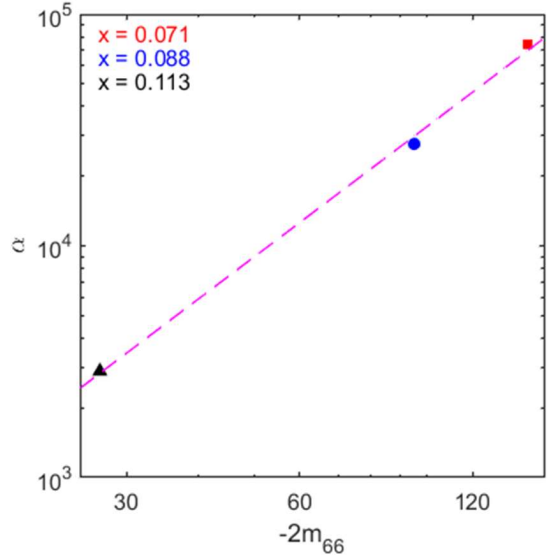

**Fig. S7 Power law fit of the relation between nematic susceptibility and sensitivity of $T_c$ to strain** Log-log plot of the normalized second derivative of $T_c$ with respect to $\epsilon_{B_{2g}}$ as a function of the elastoresistivity coefficient $-2m_{66}$ which is proportional to the nematic susceptibility $\chi_N$. The magenta dotted line is a linear fit which gives the power law $-\frac{d^2 T_c}{d\epsilon^2}\frac{1}{T_c}(\epsilon=0) \propto \chi_N^{1.9}$ .

| Doping $x$ | $C_{66}$ (GPa) | $C_{33}$ (GPa) | $C_{11}$ (GPa) | $C_{12}$ (GPa) | $C_{13}$ (GPa) | $v_{ab}$ |
|---|---|---|---|---|---|---|
| $x = 0.071$ | 22.2 | 82 | 107 | 27 | 13.5 | 0.49 |
| $x = 0.088$ | 31.4 | 79 | 111.5 | 27 | 13.5 | 0.36 |
| $x = 0.113$ | 38 | 87 | 119 | 34 | 17 | 0.32 |

**Table S1. Elastic constants and Poisson ratio for optimally and overdoped samples** All elastic constants are taken from reference (*32*). The constant $C_{13}$ was not provided in the reference but is not expected to depend heavily on doping and will be smaller than the constant $C_{12}$. We take it to be equal to half the value of $C_{12}$. The value of $v_{ab}$ is computed as described in the supplementary text.

| Doping $x$ | $p_1$ (K) | $p_2$ (K) | $p_3$ (K) |
|---|---|---|---|
| $x = 0.071$ | -865,551 | 908 | 23.4 |
| $x = 0.088$ | -305,300 | 331 | 22.2 |
| $x = 0.113$ | -21,832 | -483 | 15.2 |

**Table S2. Fit parameters for quadratic fit to the response of $T_c$ to anisotropic strain $\epsilon_{B_{2g}}$**

The parameters are defined through the fit relation $T_c = p_1 \epsilon^2 + p_2 \epsilon + p_3$. For all dopings the data is fit in the low strain region $-0.003 < \epsilon_{B_{2g}} < 0.003$.

| Material | $T_c$ (K) | $\left\lvert {}^{dT_c}\!/\!_{d\epsilon}\right\rvert$ (K) | Type of strain | Reference |
|---|---|---|---|---|
| Sn | 3.72 | 42 | Uniaxial tension along [001] | *(37)* |
| NbSe$_2$ | 7.2 | 283 | Uniaxial compression in-plane | *(35)* |
| Nb$_3$Sn | 15 | 175 | Uniaxial tension in-plane | *(36)* |
| $x$ = 0.113 Co-Ba122 | 15 | 423 | Uniaxial tension along [110] | This work |
| HgBa$_2$CuO$_{4+\delta}$ | 97 | 433 | Compression along [100], calculated from hydrostatic and uniaxial pressure | *(38)* |
| $x$ = 0.088 Co-Ba122 | 22 | 1304 | Uniaxial tension along [110] | This work |
| Sr$_2$RuO$_4$ | 1.3 | 1413 | Uniaxial compression along [100] | *(29)* |
| $x$ = 0.071 Co-Ba122 | 25 | 2672 | Uniaxial compression along [110] | This work |

**Table S3. Sensitivity to strain of various superconductors.** For each system, the value of $dT_c/d\varepsilon$ that is reported is the maximum absolute value of the derivative of the $T_c(\varepsilon)$ relation.